\documentclass[conference]{IEEEtran}
\IEEEoverridecommandlockouts
\usepackage{booktabs}
\usepackage{cite}
\usepackage{comment}
\usepackage{amsmath,amssymb,amsfonts}
\usepackage{algorithmic}
\usepackage{graphicx}
\usepackage{textcomp}
\usepackage{xcolor}
\usepackage{hyperref}
\usepackage{multirow}
\usepackage{graphicx}
\usepackage{subfig}
\usepackage[linesnumbered,ruled,vlined]{algorithm2e}
\usepackage{tikz}
\def\BibTeX{{\rm B\kern-.05em{\sc i\kern-.025em b}\kern-.08em
    T\kern-.1667em\lower.7ex\hbox{E}\kern-.125emX}}
\usepackage{balance}

\hypersetup{
  colorlinks=true,
  linkcolor=blue,
  citecolor=blue,
  filecolor=magenta,
  urlcolor=blue
}

\makeatletter
\def\ps@IEEEtitlepagestyle{%
  \def\@oddfoot{\mycopyrightnotice}%
  \def\@evenfoot{}%
}
\def\mycopyrightnotice{%
  \gdef\mycopyrightnotice{}%
}
\makeatother

\usepackage{eso-pic}
\newcommand\AtPageUpperMyleft[1]{\AtPageUpperLeft{
 \put(\LenToUnit{1cm},\LenToUnit{-1cm}){ 
     \parbox{0.5\textwidth}{\raggedright\fontsize{9}{11}\selectfont #1}} 
 }}
\newcommand{\conf}[1]{
\AddToShipoutPictureBG*{
\AtPageUpperMyleft{#1}
}
}

\def\BibTeX{{\rm B\kern-.05em{\sc i\kern-.025em b}\kern-.08em
    T\kern-.1667em\lower.7ex\hbox{E}\kern-.125emX}}
\begin{document}

\title{Design and Analysis of a Vanadium Dioxide-Based Ultra-Broadband Terahertz Metamaterial Absorber\\
}
\conf{This work has been submitted to the IEEE for possible publication. Copyright may be transferred without notice, after which this version may no longer be accessible.}


\author{\IEEEauthorblockN{Robiul Hasan \IEEEauthorrefmark{1}, and Nafisa Anjum \IEEEauthorrefmark{2}}
\IEEEauthorblockA{\IEEEauthorrefmark{1}\IEEEauthorrefmark{2}Dept. of Electrical \& Electronic Engineering (EEE), \\Rajshahi University of Engineering \& Technology, Bangladesh}
\IEEEauthorblockA{Emails: robiulhasan528527@gmail.com, nafisaanjum9999@gmail.com}
}


\maketitle

\begin{abstract}

This paper presents a VO\textsubscript{2}-based metamaterial absorber optimized for ultra-broadband, polarization-insensitive performance in the terahertz (THz) frequency range. The absorber consists of a patterned VO\textsubscript{2} metasurface, a low-loss MF\textsubscript{2} dielectric spacer, and a gold ground plane. Exploiting the phase transition of VO\textsubscript{2}, the design enables dynamic control of electromagnetic absorption. Full-wave simulations show an average absorptance of 98.15\% across a 5.38~THz bandwidth (5.72--11.11~THz) and over 99\% absorption sustained across 3.35~THz. The absorber maintains stable performance for varying polarization angles and both TE and TM modes under oblique incidence. Impedance analysis confirms strong matching to free space, reducing reflection and eliminating transmission. Parametric analysis investigates the influence of VO\textsubscript{2} conductivity, MF\textsubscript{2} thickness, and unit cell periodicity on performance. Compared to recent THz metamaterial absorbers, the proposed design achieves broader bandwidth, higher efficiency, and simpler implementation. These characteristics make it suitable for THz sensing, imaging, wireless communication, and adaptive photonic systems, and position it as a promising platform for tunable and reconfigurable THz modules.
\end{abstract}


\section{Introduction}

Metamaterials are engineered composite structures designed to possess electromagnetic properties absent in natural materials. Their periodic sub-wavelength resonators allow precise control of electromagnetic waves, enabling the development of devices such as filters, lenses, rotators, and high-efficiency absorbers. \cite{landy2008perfect}.

A key development was the metamaterial absorber by Landy et al., achieving near-unity microwave absorption through impedance-matched subwavelength structures. \cite{landy2008perfect}. Shortly after, similar principles were applied to the terahertz (THz) spectrum. Tao et al. demonstrated that a single-unit-cell THz metamaterial absorber could reach strong absorption near 1.3 THz, confirming the applicability of these concepts at higher frequencies \cite{tao2008metamaterial}. Unlike bulky conventional absorbers, metamaterial absorbers are thin and designable, making them well suited for THz systems such as sensors, imaging, detection, and stealth
\cite{tao2008metamaterial, elkorany2023design}.

Initial metamaterial absorbers were limited to narrowband responses, as they relied on a single or discrete set of resonance modes \cite{landy2008perfect, tao2008metamaterial}. To overcome this, researchers began designing absorbers with multiple resonators or layered geometries. This approach allowed distinct absorption peaks to overlap and merge into continuous broadband operation \cite{su2015terahertz, yao2016dual, qian2021polarization}.Recent work shows that single-layer graphene metasurfaces can achieve broadband, angle-tolerant, and polarization-insensitive absorption via plasmonic resonances. \cite{qian2021polarization, zhang2024graphene, ri2024tunable}.

However, passive designs lack tunability, as their spectral response is fixed after fabrication. To address this, researchers have employed tunable materials such as graphene, temperature-sensitive compounds, and electrically driven elements to allow dynamic and real-time tuning of absorption performance.

Among the various tunable materials, vanadium dioxide (VO\textsubscript{2}) is particularly promising. It undergoes a sharp and reversible phase transition near 340 K, switching from an insulating to a metallic state with a substantial change in electrical conductivity \cite{jiao2020tunable, wu2021ultra, zhang2022ultra}. This property allows VO\textsubscript{2}-based metamaterials to operate as switchable absorbers that can respond to thermal or optical inputs \cite{ren2023switchable, chen2021switchable}. In the THz range, these transitions make VO\textsubscript{2} an effective material for dynamic control of absorption states, enabling devices that can toggle between low and high absorption depending on external stimuli \cite{wu2021ultra, chen2021switchable}. Recent work on VO\textsubscript{2}-based absorbers includes narrowband, broadband, and multi-state designs using varied geometries, materials, and multilayer structures to boost bandwidth and tunability. \cite{zhang2022ultra, wang2022terahertz, jiao2020tunable}. These studies form the basis for pursuing designs that combine strong absorption, wide bandwidth, and robust polarization insensitivity. In THz absorbers, low-loss dielectrics like MgF\textsubscript{2} are often employed as spacer layers to facilitate strong field confinement and impedance matching \cite{su2015terahertz}.

This work presents a theoretical model for a VO\textsubscript{2}-based ultra-broadband THz absorber with polarization stability. The design combines a patterned VO\textsubscript{2} metasurface, an MF\textsubscript{2} spacer, and a gold ground plane to overcome bandwidth and tunability limits in earlier designs.

\section{Literature Review}

The development of THz metamaterial absorbers began in 2008 with their first reported demonstrations \cite{tao2008metamaterial}, followed by efforts to broaden the absorption bandwidth. Early designs focused on combining multiple resonant elements of different geometries within a single unit cell or stacking multiple layers to produce overlapping absorption peaks over a wider spectral range \cite{su2015terahertz, yao2016dual}. These techniques proved effective in extending the operational frequency of absorbers beyond narrowband limits.

Pan et al. (2016) designed concentric circular ring resonators supporting multiple modes, achieving 1.076 THz continuous absorption by spatially distributing resonances across the band. \cite{pan2016broadband}. Su et al. (2015) demonstrated a dual-band THz absorber using multilayer graphene and MgF\textsubscript{2} films on gold, achieving polarization-insensitive, near-perfect absorption at 4.95 THz and 9.2 THz through electric and magnetic dipole resonances. The stacked structure also introduced hyperbolic dispersion for more flexible spectral design \cite{su2015terahertz}. Rahmanzadeh et al. (2018) extended the multilayer graphene approach using a designed metasurface, achieving polarization-insensitive, ultra-broadband THz absorption \cite{rahmanzadeh2018multilayer}. This marked a shift from single-resonance designs to multi-resonant broadband absorbers. Recent studies show that well-designed single-layer metasurfaces can achieve broadband absorption. Qian et al. (2021) used circular graphene patches of two radii to realize over 90\% absorption from 1.34 to 3.36 THz, with polarization and angle tolerance \cite{qian2021polarization, zhang2024graphene}. These efforts highlight the shift from narrowband to broadband absorbers via geometric and material optimization.

Despite these advancements, passive broadband absorbers have a fixed spectral response once fabricated \cite{qian2021polarization, su2015terahertz}. This limitation has led to a growing interest in tunable or reconfigurable metamaterials. By incorporating active materials, absorbers can adapt their performance based on outside factors, including temperature, voltage, or optical excitation\cite{zhang2024graphene}. Within the class of tunable materials, VO$_2$ stands out due to its ability to transition from insulating to metallic behavior near 340~K, causing a large change in its electrical conductivity.\cite{jiao2020tunable, wu2021ultra}. This behavior enables dynamic tuning of the absorber’s electromagnetic properties in the THz range. Jiao et al. (2020) introduced a dual-broadband absorber using elliptical ring resonators made of VO\textsubscript{2} \cite{jiao2020tunable}. In the metallic phase, the structure showed two broadband absorption bands above 90\% at 2.32 and 2.03 THz. Similarly, Huang et al. (2020) used four VO\textsubscript{2} loops on a dielectric–metal stack to achieve a single 2.45 THz-wide absorption band. \cite{huang2020broadband}. These studies demonstrate the effectiveness of VO\textsubscript{2} in expanding the tunability and functionality of THz absorbers without requiring complex biasing circuits or active electronics \cite{jiao2020tunable, huang2020broadband, chen2021switchable}.

MgF\textsubscript{2} has been widely used in THz absorbers as a low-loss, low-index dielectric spacer or capping layer. Its low real permittivity improves impedance matching with free space by increasing the dielectric's intrinsic impedance, thus reducing reflection. These properties have enabled broadband, polarization-insensitive absorption in several designs, such as the MgF\textsubscript{2}/Cr multilayer absorber reported by Chen et al., which achieved over 90\% absorption from 900 to 1900 nm. \cite{chen2021ultra}.

In summary, the literature reveals steady progress from narrowband metamaterial absorbers to broadband and tunable designs. Techniques such as multi-resonator geometries, multilayer substrates, and the integration of phase-change materials like VO\textsubscript{2} have expanded the design space of THz absorbers \cite{zhang2022ultra, wu2021ultra, ren2023switchable}. These insights provide the foundation for the proposed VO\textsubscript{2}-based absorber in this work, which aims to achieve ultra-broadband absorption with stable polarization response and potential thermal tunability.

\section{Methodology}

\begin{table}[htbp]
\caption{Structural Dimensions of the Absorber}
\centering
\renewcommand{\arraystretch}{1.2}
\setlength{\tabcolsep}{3pt}
\begin{tabular}{|c|c|c|c|c|c|c|c|c|c|}
\hline
\textbf{Parameter} & $p$ & $t_{\text{VO}_2}$ & $t_d$ & $t_m$ & $R_{o1}$ & $R_{i1}$ & $w$ & $R_{o2}$ & $R_{i2}$ \\
\hline
\textbf{Value ($\mu$m)} & 13 & 0.2 & 6 & 0.02 & 6.5 & 5.5 & 1 & 2 & 1 \\
\hline
\end{tabular}
\label{tab:parameters}
\end{table}

\begin{figure}[htbp]
\centerline{\includegraphics[width=\linewidth]{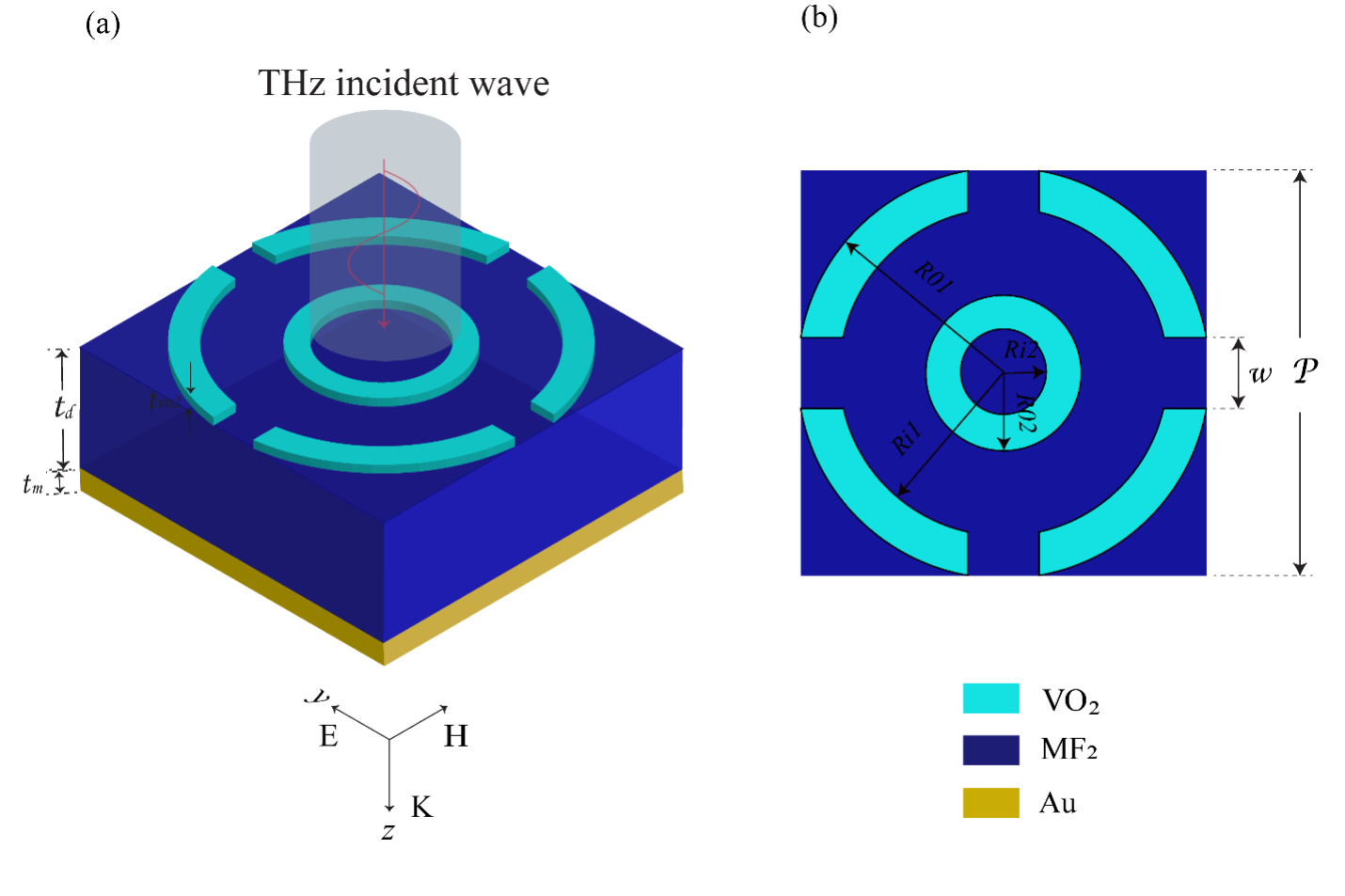}}
\caption{Structural Configuration of the Designed Absorber: (a) 3D schematic showing layer stack and incident THz wave direction; (b) top view with labeled geometrical parameters.}
\label{fig:model}
\end{figure}

\begin{figure*}[t]
\centering
\subfloat[Absorption and reflection spectra.]{\includegraphics[width=0.29\linewidth]{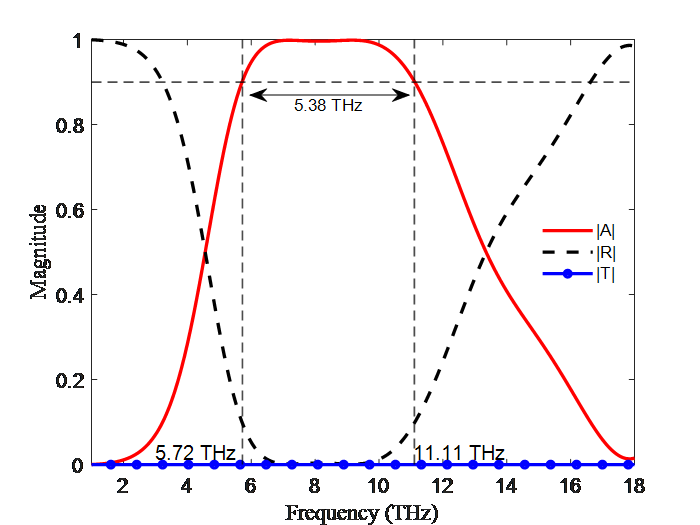}} \hfill
\subfloat[Absorbance for different conductivity of VO\textsubscript{2}.]{\includegraphics[width=0.33\linewidth]{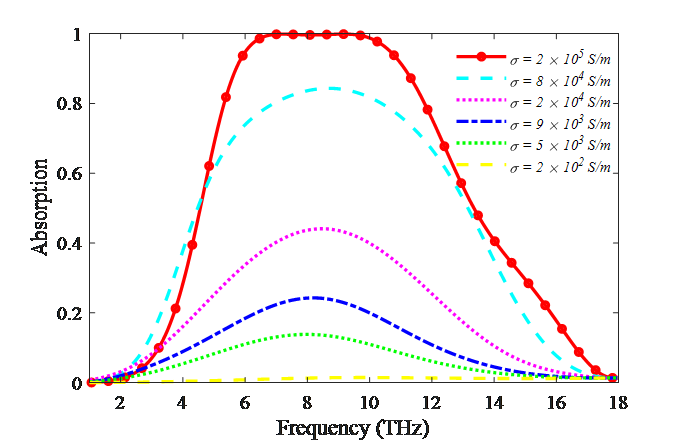}} \hfill
\subfloat[Absorption for variation of t\textsubscript{VO\textsubscript{2}}.]{\includegraphics[width=0.31\linewidth]{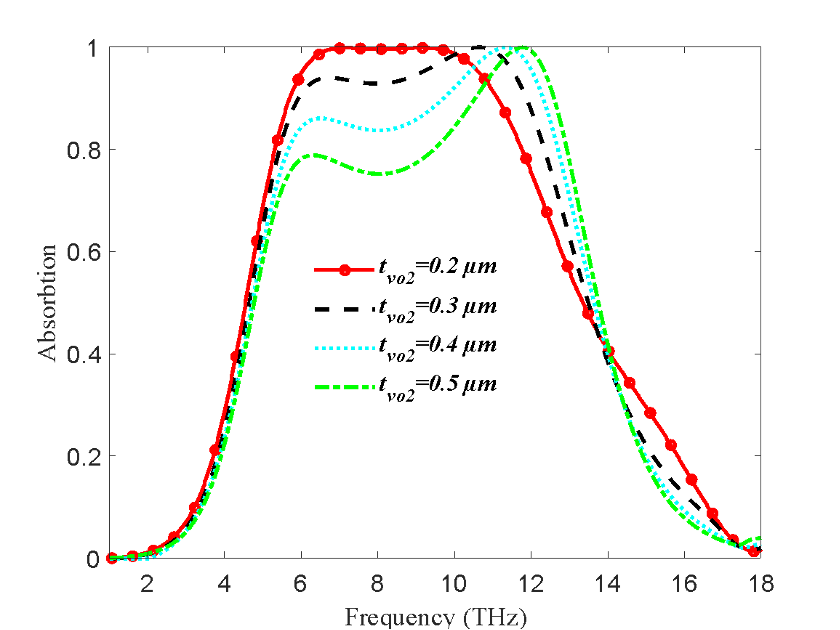}} \\
\subfloat[Absorption for different periodicities]{\includegraphics[width=0.31\linewidth]{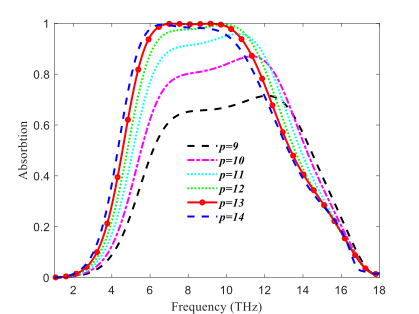}} \hfill
\subfloat[Absorption for variation of t\textsubscript{d}]{\includegraphics[width=0.32\linewidth]{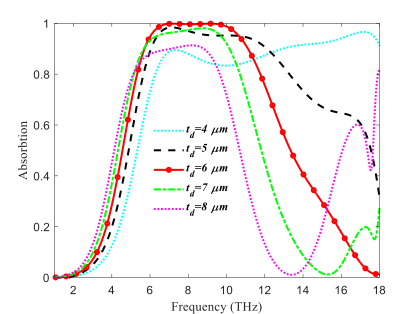}} \hfill
\subfloat[Normalized impedance (Z\textsubscript{r})]{\includegraphics[width=0.31\linewidth]{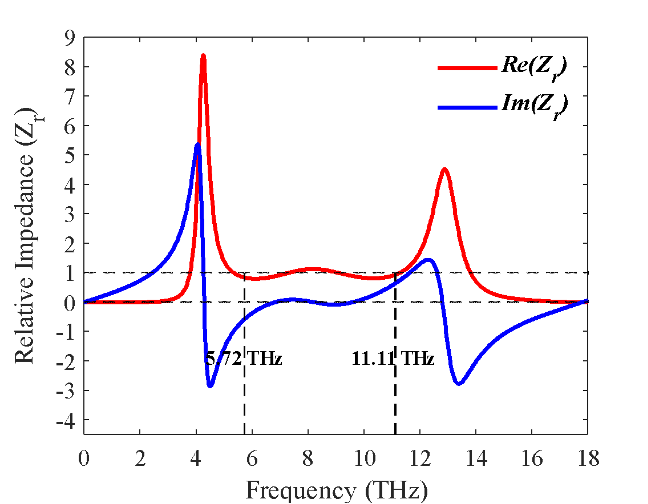}}

\caption{Performance evaluation showing absorption and reflection spectra, tunability with VO\textsubscript{2} properties, effects of periodicity and dielectric spacing, and broadband impedance matching.}
\label{fig:simulation}
\end{figure*}

The designed metamaterial absorber utilizes a three-layer configuration tailored for broadband absorption in the terahertz range. As presented in Fig.~\ref{fig:model}, the top layer includes a patterned VO\textsubscript{2} metasurface, featuring a central closed circular ring and two outer concentric split-ring resonators (SRRs). This configuration supports multiple resonance modes and enables strong coupling, facilitating wideband absorption. The thickness of the VO\textsubscript{2} layer is set to $t_{\text{VO}_2} = 0.2~\mu$m. Beneath the metasurface lies a dielectric spacer made of MF\textsubscript{2}, which acts as a Fabry–Pérot-like cavity. Selected for its low dielectric loss and relative permittivity of $\varepsilon_r = 1.9$, MF\textsubscript{2} assists in impedance matching between free space and the absorber. The thickness of this layer is set to $t_{d}=6~\mu\text{m}$. The substrate is backed by a gold (Au) film having a thickness of $t_{m}=0.02~\mu\text{m}$ and an electrical conductivity of $\sigma=4.56\times10^{7}~\text{S/m}$.Gold is a great choice for the bottom layer because it conducts electricity well in the THz range, has less scattering of free electrons, and experiences lower energy loss. Additionally, gold is very stable and doesn’t rust easily, unlike many other metals, making it reliable in different conditions. This metallic layer eliminates transmission and reflects residual energy.

In the terahertz (THz) range, the optical behavior of VO$_2$ is often represented by the Drude model, which defines its complex permittivity as
 \cite{hossain2024multi}:
\begin{equation}
\varepsilon(\omega) = \varepsilon_{\infty} - \frac{\omega_p^2}{\omega^2 + i\gamma\omega}
\label{eq:drude}
\end{equation}
Here, $\varepsilon_{\infty} = 12$ denotes the high-frequency dielectric permittivity, $\gamma = 5.75 \times 10^{13}~\text{rad/s}$ represents the collision frequency, $\omega$ is the angular frequency, and $\omega_p$ corresponds to the plasma frequency. This formulation describes the frequency-dependent behavior of VO\textsubscript{2} and its associated absorption properties.

The variation of plasma frequency $\omega_p$ with respect to electrical conductivity $\sigma$ is described by the following expression
\cite{hossain2024multi}:
\begin{equation}
\omega_p^2(\sigma) = \frac{\sigma}{\sigma_0} \cdot \omega_p^2(\sigma_0)
\label{eq:plasma}
\end{equation}

In this study, the reference conductivity is assigned as $\sigma_0 = 3 \times 10^{5}~\text{S/m}$, corresponding to a plasma frequency of $\omega_p(\sigma_0) = 1.4 \times 10^{15}~\text{rad/s}$~\cite{zhang2022ultra}. The electrical conductivity of VO$_2$ is considered to vary from 200~S/m in its insulating phase up to $2 \times 10^{5}~\text{S/m}$ when metallic~\cite{jiao2020tunable}. To analyze the absorber’s reflection and absorption properties, a finite element method (FEM) simulation with periodic boundary conditions is performed, as shown in Fig.~\ref{fig:model}(a). The unit cell is periodically extended along the $x$ and $y$ directions, while a perfectly matched layer (PML) is applied in the $z$ direction to minimize reflections of outgoing waves.

Due to the gold backing layer being considerably thicker than the skin depth, transmission of electromagnetic waves is effectively suppressed. As a result, the transmittance $T(\omega)$ is taken as zero, and the absorptance $A(\omega)$ is determined by \cite{chen2021switchable}:

\begin{equation}
A(\omega) = 1 - R(\omega) = 1 - \left|S_{11}(\omega)\right|^2
\end{equation}

Here, $R(\omega)$ represents the reflectance, while $S_{11}(\omega)$ corresponds to the reflection coefficient derived from the S-parameters.

\section{Results and Discussion}

\begin{figure*}[t]
\centering

\subfloat[Polarization angle effect]{\includegraphics[width=0.30\linewidth]{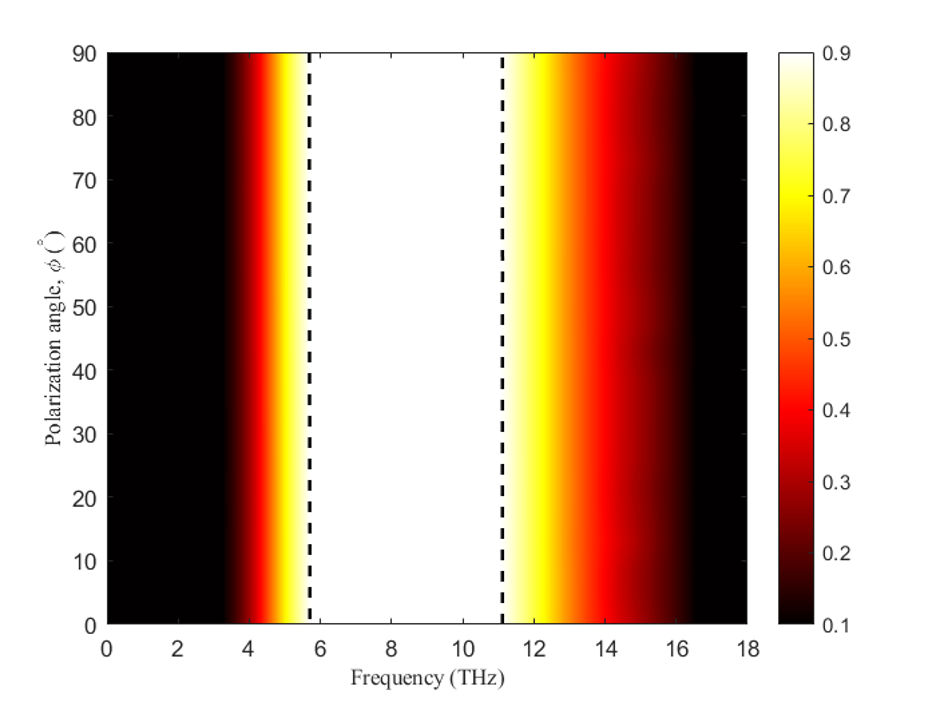}} \hfill
\subfloat[TE-mode incidence response]{\includegraphics[width=0.30\linewidth]{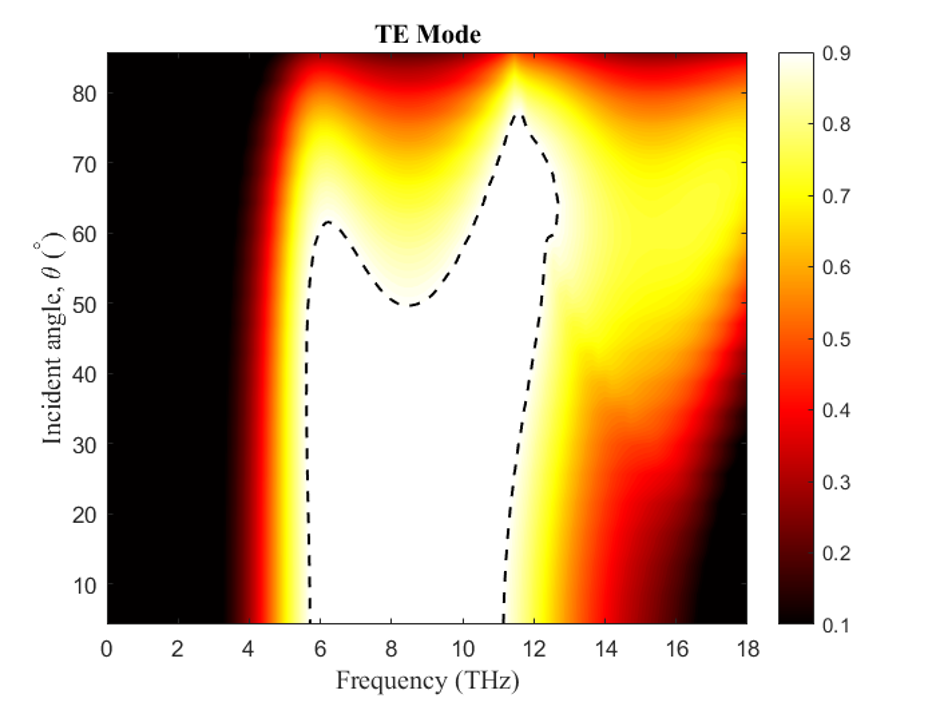}} \hfill
\subfloat[TM-mode incidence response ]{\includegraphics[width=0.30\linewidth]{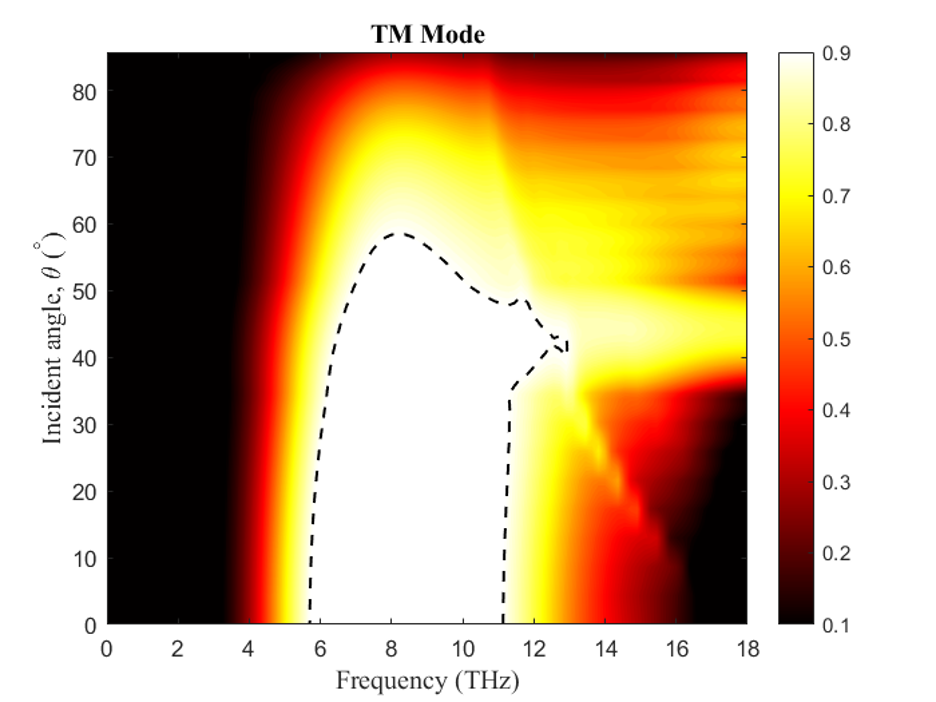}}

\caption{Absorption response under varying polarization angles and incident modes (TE and TM), demonstrating polarization-insensitive and angle-stable performance.}
\label{fig:simulation_results}
\end{figure*}

This section presents the simulation outcomes and performance evaluation of the proposed VO\textsubscript{2}-based metamaterial absorber. All analyses were performed using frequency-domain simulations under various incident conditions to understand the absorber’s electromagnetic behavior. We present simulation results that assess the absorber's broadband response, polarization insensitivity, and impedance characteristics. The design's performance is also compared with existing VO\textsubscript{2}-based absorbers to highlight its improvements in bandwidth and tunability.

\subsection{Simulation Results}\label{AA}

Fig.~\ref{fig:simulation}(a) presents the simulated absorption ($|A|$), reflection ($|R|$), and transmission ($|T|$) responses of the proposed absorber over the 0.1--18~THz frequency range. The structure achieves a wide absorption band from 5.72~THz to 11.11~THz, corresponding to a 5.38~THz bandwidth. Within this region, a stable absorption exceeding 99\% is sustained between 6.57~THz and 9.92~THz, covering a span of 3.35~THz. The near-zero transmission across the spectrum highlights the role of the gold ground plane in eliminating energy leakage, confirming that the high absorption primarily results from suppressed reflection. These outcomes demonstrate the absorber's capability for broadband, high-efficiency operation in the terahertz regime.

As depicted in Fig.~\ref{fig:simulation}(b), the absorption performance of the designed metamaterial absorber is analyzed for different VO$_2$ conductivities, varying between 200~S/m and $2 \times 10^{5}$~S/m.Higher conductivity values lead to notable improvements in both bandwidth and peak absorption. At $2 \times 10^5$~S/m, the design attains near-unity absorption over a wide terahertz range, verifying the phase transition of VO\textsubscript{2} from an insulating to a metallic state. This controllable behavior highlights the absorber’s suitability for adaptive electromagnetic functionality in dynamically changing conditions.

Fig.~\ref{fig:simulation}(c) presents the absorption characteristics of the proposed absorber for VO\textsubscript{2} layer thicknesses between 0.2 and $0.5~\mu\text{m}$. The results show that high absorption efficiency is maintained across the broadband range for all cases, with the best performance occurring at $t_{\text{VO}_2} = 0.2~\mu\text{m}$, where the structure achieves near-unity absorption over the broadest frequency range.

Fig.~\ref{fig:simulation}(d) illustrates the absorption response of the proposed absorber for various periodicities (\( p \)) of the unit cell. Variations in \( p \) lead to noticeable shifts in the resonance frequencies, which in turn influence both the position and shape of the absorption spectrum. Smaller or larger values of \( p \) can slightly degrade the broadband performance by altering the coupling between resonant elements. Selecting an optimal periodicity allows for better mode overlap, resulting in stronger absorption peaks and a wider operational bandwidth within the terahertz range.

In Fig.~\ref{fig:simulation}(e), the absorption spectra are shown for various dielectric spacer thicknesses \( t_d \) ranging from \(4~\mu\text{m}\) to \(8~\mu\text{m}\). The results demonstrate that dielectric thickness has a significant influence on both the bandwidth and peak absorptance of the absorber. The optimal performance is observed at \( t_d = 6~\mu\text{m} \), where the structure maintains near-unity absorption across a broad frequency range. This enhancement is primarily due to the Fabry–Pérot-like cavity effect created within the dielectric spacer, which improves impedance matching between the absorber and free space. Deviations from this optimal thickness, whether thinner or thicker, shift the resonance frequencies and reduce the constructive overlap of multiple modes, leading to a narrower bandwidth and decreased absorption efficiency. These findings highlight the critical role of accurately selecting \( t_d \) to achieve high-efficiency, broadband operation in the terahertz spectrum.

Fig.~\ref{fig:simulation}(f) shows the normalized impedance \( Z_r \) of the absorber, defined as the ratio of the structure’s effective impedance to that of free space. Efficient absorption occurs when the metamaterial impedance matches its surrounding medium, minimizing reflection. The results indicate that near-unity absorption is obtained when the real part of \( Z_r \) is close to 1 and the imaginary part is near 0, confirming excellent impedance matching. Under these conditions, incident THz waves are predominantly absorbed rather than reflected or transmitted. The impedance-matched profile over a broad frequency band highlights the absorber’s capability for broadband, high-efficiency energy dissipation.

Fig.~\ref{fig:simulation_results}(a) depicts the absorber’s Response observed for polarization angles within the range of \(0^\circ\) and \(90^\circ\). The results describe that absorption consistently exceeds 90\% throughout the operational bandwidth, verifying the polarization-independent behavior of the design. This stability arises from the symmetric configuration of the VO\textsubscript{2} metasurface elements, which ensures uniform electromagnetic response regardless of the incident field’s polarization direction. Such performance is particularly important for practical terahertz applications, where the polarization state may vary dynamically or be unknown in advance.

\begin{table*}[ht]
\centering
\caption{Comparative Analysis with Existing Terahertz Absorber Designs}
\resizebox{\textwidth}{!}{
\begin{tabular}{|p{3.5cm}|c|c|p{6cm}|}
\hline
\textbf{Reference} & \textbf{Absorption} & \textbf{Bandwidth (THz)} & \textbf{Material and Design Approach} \\
\hline
Wu \textit{et al.}, 2021 \cite{wu2021ultra} & $>90$\% & 3.30 & VO\textsubscript{2}-based, Broadband Tunable Absorber \\
\hline
Xie \textit{et al.}, 2021 \cite{xie2021tunable} & $>90$\% & 2.40 & Graphene‑based, Polarization‑Insensitive, Tunable via Fermi Level \\
\hline
Wang \textit{et al.}, 2022 \cite{wang2022terahertz} & $>90$\% & 4.26 & VO\textsubscript{2}-based, Multi-Ring Structure, Broadband Tunability \\
\hline
Elkorany \textit{et al.}, 2023 \cite{elkorany2023design} & 100\% & 0.25 & Metallic, Polarization-Insensitive, Narrowband Design \\
\hline
Ren \textit{et al.}, 2023 \cite{ren2023switchable} & $>90$\% & 5.08 & VO\textsubscript{2}-based, Switchable (Ultra-Broadband/Dual-Band) Absorber \\
\hline
Ri \textit{et al.}, 2024 \cite{ri2024tunable} & $\sim$90\% & 2.18 & Graphene-based, Complementary Split-Ring, Highly Tunable \\
\hline
Zhang \textit{et al.}, 2024 \cite{zhang2024graphene} & $\sim$92\% & $\sim$2.5 & Graphene-based, Dual-Broadband Absorber, Independent Tunability \\
\hline
Xiong \textit{et al.}, 2025 \cite{xiong2025terahertz} & $>95$\% & 4.07 & VO\textsubscript{2}\,and\,Graphene Hybrid, Independently Tunable Absorbance \\
\hline
Ge \textit{et al.}, 2025 \cite{ge2025dual} & $>90$\% & 1.93  & VO\textsubscript{2}-based, Microfluidic Square Resonator, Dual-Function Absorber and Sensor, Polarization-Insensitive \\
\hline
Ryu \textit{et al.}, 2025 \cite{ryu2025theoretical} & $>90$\% & 4.70  & VO\textsubscript{2}-based, Single Resonator, Ultra-Broadband, Polarization-Insensitive, High Modulation Depth \\
\hline
\textbf{This Work} & \textbf{$>$90\%} & \textbf{5.38} & \textbf{VO\textsubscript{2}-based, Polarization-Insensitive, Ultra-Broadband, High-Efficiency} \\
 & \textbf{$>$99\%} & \textbf{3.35} &  \\
\hline
\end{tabular}}
\label{table:comparison}
\end{table*}

Fig.~\ref{fig:simulation_results}(b) and Fig.~\ref{fig:simulation_results}(c) illustrate the absorption characteristics of the designed absorber for various incident angles in TE and TM polarization modes, respectively. Specifically, Figure~3(b) analyzes the TE-polarized response under multiple angles of incidence.The structure exhibits stable absorption at oblique incidence, supporting angular robustness along with polarization independence. Similarly, in the TM mode shown in Fig.~\ref{fig:simulation_results}(c), the broadband absorption remains effective over a comparable frequency range and is less sensitive to angular variations, especially below 60°. These findings verify that the absorber maintains stable performance across various angles and polarizations.

\subsection{Comparative Study}

For performance assessment the effectiveness of the proposed VO\textsubscript{2}-based absorber, we compare its performance with several recent terahertz absorber designs reported between 2021 and 2025, as summarized in Table~\ref{table:comparison}.

Wu \textit{et al.}~\cite{wu2021ultra} introduced a VO\textsubscript{2}-based broadband absorber achieving over 90\% absorption with a 3.30~THz bandwidth. Similarly, Wang \textit{et al.}~\cite{wang2022terahertz} proposed a multi-ring VO\textsubscript{2} configuration covering 4.26~THz, while Ren \textit{et al.}~\cite{ren2023switchable} designed a switchable absorber with 5.08~THz bandwidth. Despite their broadband characteristics, none of these designs surpass our 5.38~THz range. More recent efforts include Xiong \textit{et al.}~\cite{xiong2025terahertz}, who employed a VO\textsubscript{2}-graphene hybrid to achieve 4.07~THz, and Ryu \textit{et al.}~\cite{ryu2025theoretical}, whose single-resonator design provided 4.70~THz coverage with strong modulation capabilities. In contrast, graphene-based absorbers have generally achieved narrower bandwidths. For example, Xie \textit{et al.}~\cite{xie2021tunable} reported a polarization-insensitive design with a 2.40~THz range, while Zhang \textit{et al.}~\cite{zhang2024graphene} demonstrated independent tunability over dual bands spanning approximately 2.5~THz. Ri \textit{et al.}~\cite{ri2024tunable} utilized a complementary split-ring configuration, offering 2.18~THz of tunable absorption. Although these designs showcase flexibility, they fall short in bandwidth and absorption uniformity. Elkorany \textit{et al.}~\cite{elkorany2023design} proposed a metallic absorber with perfect absorption but only a narrow 0.25~THz bandwidth. Ge \textit{et al.}~\cite{ge2025dual} introduced a microfluidic VO\textsubscript{2} structure that doubles as a sensor, achieving 1.93~THz bandwidth, which is significantly lower than our model. Overall, our absorber delivers superior performance by maintaining 90\% absorption over a broader 5.38~THz range while also ensuring polarization insensitivity and structural simplicity. These advantages make it a compelling candidate for practical THz applications in sensing, filtering, and imaging.

\section{Conclusion}

This paper presented a high-performance VO\textsubscript{2}-based metamaterial absorber engineered for broadband and polarization-insensitive operation in the terahertz region. The design attained a broad absorption bandwidth of 5.38~THz (ranging from 5.72~THz to 11.11~THz) with an average absorptance of approximately 90\%. It maintained strong performance across varying polarization angles and incident modes, while also demonstrating tunability through the phase transition properties of VO\textsubscript{2}. Simulation results confirmed excellent impedance matching and minimal reflection, enabling near-unity absorption without transmission leakage. These findings contribute to ongoing efforts to develop efficient and adaptable THz absorbers for use in sensing, imaging, and electromagnetic shielding. The ability to dynamically tune the absorption response using VO\textsubscript{2} highlights its potential for real-time, reconfigurable terahertz applications. The main contribution of this work is a theoretically validated absorber design that offers a rare combination of ultra-broadband coverage, high absorptance, polarization stability, and tunability, all within a structurally simple three-layer configuration. Future research may extend this design toward experimental fabrication, thermal or electrical switching integration, and the development of hybrid absorbers combining VO\textsubscript{2} with other tunable materials. We encourage further exploration of VO\textsubscript{2}-based architectures for building compact, programmable photonic devices, and recommend validating the proposed absorber under real-world operating conditions to enable next-generation THz system deployment.
\bibliography{references}

\end{document}